\documentclass[twocolumn,showpacs,preprintnumbers,amsmath,amssymb]{revtex4-1}
\usepackage{graphicx}
\usepackage{dcolumn}
\usepackage{bm}
\usepackage{color}
\begin{document}
%
%
\title{Experimental evidence that electrical fatigue failure obeys a generalized Coffin-Manson law}
\author{Xiangtong He and John Y. Fu}
\email{johnyfu2016@gmail.com}
%
%
\affiliation{Department of Mechanical and Aerospace Engineering, The State University of New York, Buffalo, NY 14260, USA}
%
%
%
%
\begin{abstract}
The empirical Coffin-Manson law has been used to characterize the low-cycle mechanical fatigue failure of metallic materials for decades. Our experimental studies reported in this letter have shown that the electrical fatigue failure in dielectrics can be well described by a fitting function having the same mathematical expression as that of the Coffin-Manson law. This observation indicates that the physical mechanism beneath the formation and evolution of atomic disordered structures, the key factor influencing both mechanical and electrical fatigue, might be the same.
\end{abstract}
%
%
%
%
\maketitle
%
%
%
%
\section{Introduction}
Electrical fatigue or polarization fatigue phenomena in ferroelectric materials have been extensively investigated in the past decades. However, a general understanding of the origin of such dielectric deterioration is still lacked \cite{tagantsev2001,lou2009}. In this letter, our experimental studies on polarization fatigue failure of poly(vinylidene fluoride-trifluoroethylene) [P(VDF-TrFE) 75/25mol\%] copolymer films are reported. The objective of our investigation is to attempt to tackle the polarization fatigue problem from a different perspective and gain a better understanding of it. To be specific, we would like to test whether electrical fatigue can be described by a certain mathematical formula of mechanical fatigue law or not. Our approach is based on the reasoning and argument given below.

When we apply an external voltage to a piece of insulating material, it will undergo volume deformation due to electromechanical coupling effects (electrostriction and flexoelectricity for all dielectrics and plus piezoelectricity for ferroelectric materials). If the applied voltage is a periodic signal with a long duration (electrical cyclic loading), the deformation of the dielectric will become a cyclic process. Thus, the electrical cyclic loading for dielectrics is equivalent to the mechanical cyclic loading for metallic materials. Certainly, we would conjecture that, from the viewpoint of equilibrium thermodynamics, there must be some resemblances between mechanical and electrical fatigue.

Furthermore, if the above mentioned deformation remains in the elastic range, there is a pair of balanced forces, $\nabla_{r}W_{e}=\nabla_{r}W_{m}$, inside the dielectric; here $W_{e}$ is the electrostatic energy stored in the dielectric, $\nabla_{r}$ is the mathematical symbol representing the gradient with respect to the direction $r$, and $W_{m}$ is the mechanical energy, due to the induced deformation, stored in the dielectric. In order to prevent partial discharge channels or, more generally, electrical-breakdown structure precursors, which would eventually lead to the occurrence of electrical breakdown in the dielectric, from initiating and growing, the following force inequality must be satisfied \cite{zeller1984}: $\nabla_{r}W_{e}\leq\nabla_{r}W_{max}$; here $W_{max}$ represents the maximum mechanical energy allowed to be stored beyond which electrical-breakdown structure precursors will initiate and grow. In this case, the initiating and growing process of electrical-breakdown structure precursors can be regarded as that of microcracks in mechanical fatigue failure. Thus, there must also be some resemblances between mechanical and electrical fatigue failure.

In this letter, we attempt to test if the electrical fatigue failure in dielectrics can be fitted by using a function that has the same mathematical expression as that of the Coffin-Manson law. Before introducing our fitting function, we first write down the Coffin-Manson law as follows.

The low-cycle fatigue of metallic materials is described by the Coffin-Manson law \cite{coffin1954,manson1953}, which is given below.
\begin{equation}
\frac{\epsilon_{p}}{2}\approx\epsilon_{f}(2N_{f})^{-\beta_{CM}},
\label{coffinmanson}
\end{equation}
here $\epsilon_{p}$ is the induced plastic strain; $\epsilon_{f}$ is an empirical value known as the fatigue ductility coefficient that represents the failure strain for a single reversal; $N_{f}$ is the number of reversals to failure (life cycles) and $\beta_{CM}$ is called the Coffin-Manson exponent.

Now we substitute $\frac{\epsilon_{p}}{2}$ in the above equation with $P_{r}$ and $\epsilon_{f}(2)^{-\beta_{CM}}$ with $P_{f}$; then we have the following fitting function
\begin{equation}
P_{r}\approx P_{f}(N_{f})^{-\beta_{CM}},
\label{fitting-function}
\end{equation}
where $P_{r}$ is the remnant polarization; $P_{f}$ is defined as the electrical failure coefficient that is the failure polarization (the remnant polarization just before electrical breakdown occurs) for a single reversal. For simplicity we define Eq.~(\ref{fitting-function}) as a generalized Coffin-Manson law.

To show the potential merit of our studies, it might be necessary to consider and answer two questions before we present our experimental results. The first question is why we choose the Coffin-Manson law as a template for proposing Eq.~(\ref{fitting-function}) to describe electrical fatigue failure? Our consideration is that the correctness of the Coffin-Manson law has been experimentally verified during the past six decades, and moreover, the Coffin-Manson law is an empirical law and its physical mechanism is still not completely clear. Therefore, if we could experimentally verify that electrical fatigue failure in dielectrics can be fitted by Eq.~(\ref{fitting-function}), then our studies may provide a different route where not only can some experimental and analytical techniques in mechanical fatigue field be borrowed to investigate electrical fatigue but also the similarity between these two fatigue behavior might reveal that their physical origins at the atomic level could be the same. The second question is why we consider polarization fatigue of ferroelectric materials in our studies? In low-cycle fatigue of metallic materials where the Coffin-Manson law is applicable, the stress - strain relationship of the studied metallic sample is no longer linear (plastic deformation occurs). Similarly, in polarization fatigue of ferroelectric materials, the electric field - polarization relationship of the studied dielectric sample is also nonlinear (hysteresis loop appears). Therefore, it would be reasonable to compare polarization fatigue failure data to the fitting curve given by Eq.~(\ref{fitting-function}).

In the following sections, our material preparation and experiment procedures are introduced; then follow discussions of experimental results and concluding remarks.

\begin{figure}[h!]
\begin{center}
\includegraphics[width=1.0\columnwidth]{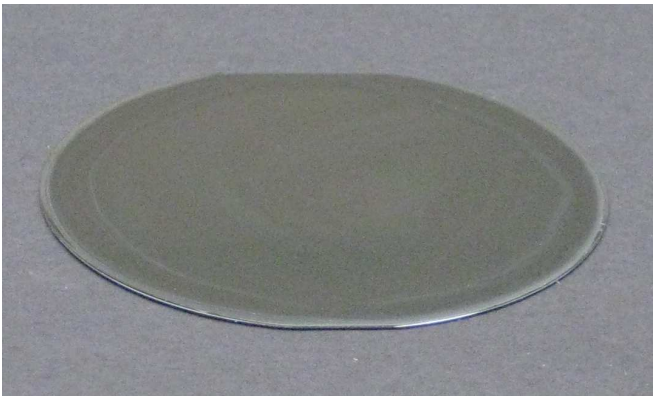}
\vspace{-0.1in}
\caption{The P(VDF-TrFE) copolymer film coated on a $\mathrm{Pt/Ti/SiO_{2}/Si(100)}$ substrate.}
\vspace{-0.15in}
\end{center}
\end{figure}

\section{\label{sec:material}Material preparation and experimental procedures}
P(VDF-TrFE) 75/25mol\% copolymer films were used in our studies. Its powder was provided by Solvay Solexis and the procedure of preparing P(VDF-TrFE) films is briefly summarized as follows (our procedure is a modified version of the one reported in Ref.~\cite{zhu2006}): (1) Dissolving P(VDF-TrFE) powder in a butanone solvent; the weight ratio of the powder to the solvent is 12:100; (2) Stirring the P(VDF-TrFE)-butanone solution by using a magnetic stir bar on the surface of a hot plate at 90$^{\circ}$C degrees for 1 hour; (3) Spin-coating a thin layer of the solution on the top of a $\mathrm{Pt/Ti/SiO_{2}/Si(100)}$ wafer, previously cleaned with alcohol/DI water, by using a Laurell spin coater. The coating time and the coating rate (RPM) vary in different situations and were used to control the thickness of the layer; (4) Placing the spin-coated wafer in a pre-heated oven at 150$^{\circ}$C degrees for 8 minutes; (5) Then turning off the power to the oven and letting the wafer cool down to room temperature naturally.

The fabricated film is shown in Fig. 1. The thickness of the film was measured and its average value is $\sim\mathrm{3\mu m}$. The Pt layer of the wafer was used as the ground electrode and top electrodes were fabricated on the surface of the film; each of top electrodes is a solid circle, covered by silver paste (PELCO colloidal silver paste from Ted Pella), with a radius of $\sim\mathrm{2 mm}$; there is an appropriate space to separate one top electrode from others so that the occurrence of electrical breakdown at that spot will not affect later measurement at other spots.

Polarization fatigue phenomena of our P(VDF-TrFE) film were studied by using a standard Sawyer-Tower circuit. The input and output waveforms of this Sawyer-Tower circuit are shown in Fig. 2; the triangle waveform is provided by a function generator (Stanford Research Systems DS345) and this signal is also applied, via a high voltage amplifier (Trek 609B), across the top and ground electrodes of the aforementioned film; the distorted waveform, shown in Fig. 2(a), represents the voltage measured across the reference capacitor of the Sawyer-Tower circuit; when electrical breakdown occurs in the film, the voltage will be totally applied across the reference capacitor so that we can see in Fig. 2(b) that the signal goes beyond the voltage limit of the oscilloscope (Agilent DSO3062A) connected to the circuit. Here we have to emphasize that, for safety reason and oscilloscope protection, a high voltage probe (Agilent 10076B 100:1 passive probe) had been used during our experiment. The observed evolution of polarization fatigue in our experiment is quite similar to that reported in Refs.~\cite{zhu2006,yuan2011} - they are all frequency dependent! This is not surprised because most P(VDF-TrFE) copolymers are viscoelastic materials at room temperature, which means that their responses to external perturbations are always time-variant. The ``{\it difference}" between our observations and theirs is that an asymmetric E-P hysteresis loop is reported in our paper (see Fig. 3) but was not mentioned in theirs. Fig. 3 shows the measured E-P hysteresis loops of our polymer film driven by a triangle voltage waveform with a frequency of 100Hz and an amplitude of $0.7$MV/cm for 12 hours; the arrow indicates the trend of the remnant polarization fatigue, an increase in hysteresis loop cycles would decrease the values of $P_{r}$. Actually, what we have observed in Fig. 3 can also be seen in both Figs. 1 of Refs.~\cite{zhu2006,yuan2011}. Our comments on the ``{\it difference}" and the asymmetric fatigue phenomenon will be given in Section~\ref{sec:results}.

In Eq.~(\ref{coffinmanson}), the fatigue ductility coefficient, $\epsilon_{f}$, of metallic materials can be determined quite straightforward: apply a large load to a piece of a certain metallic material; make sure that the load is large enough to lead to the occurrence of material failure within a single reversal; $\epsilon_{f}$ is measured as the failure strain for that material and can be regarded as an empirical constant. However, determining the electrical failure coefficient, $P_{f}$, in Eq.~(\ref{fitting-function}) is a different story. There are two problems to prevent us from determining $P_{f}$ the way we measure $\epsilon_{f}$. The technique problem is that, compared with mechanical fatigue failure, electrical fatigue failure (electrical breakdown) can occur in a very short time. It is difficult to measure $P_{f}$ just before the occurrence of electrical breakdown. The fundamental problem is that P(VDF-TrFE) is a viscoelastic material at room temperature. Thus, even if one can somehow solve the first technique problem, but there is no way that one can deal with the following situation: under applied voltage signals with different amplitudes and frequencies, the measured values of $P_{f}$ will be different too. Our strategies to determine $N_{f}$, $P_{r}$, $P_{f}$, and $\beta_{CM}$ are summarized below.

\begin{figure}[h!]
\begin{center}
\includegraphics[width=1.0\columnwidth]{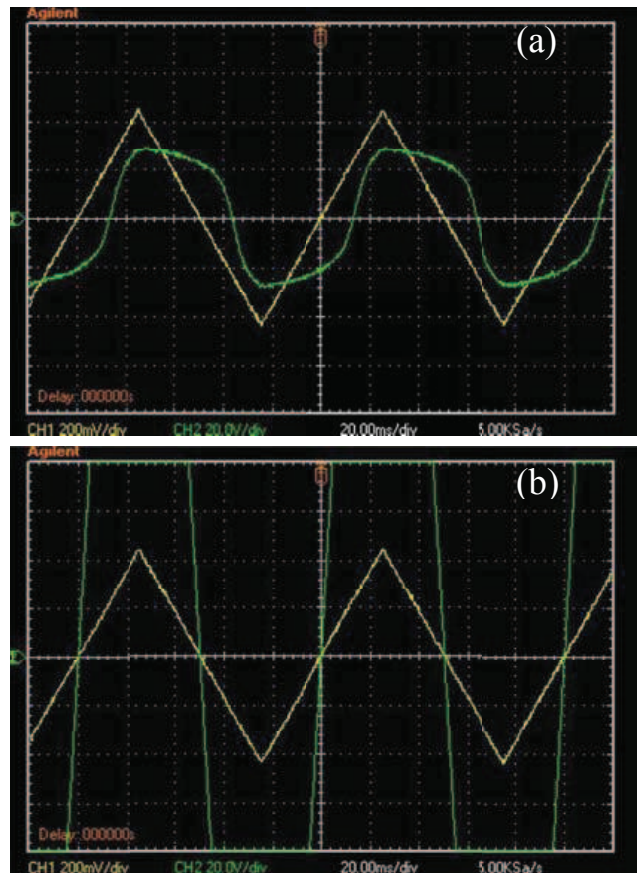}
\vspace{-0.1in}
\caption{The waveforms collected from the oscilloscope connected to a standard Sawyer-Tower circuit: (a) the triangle waveform is the input voltage signal and the distorted waveform represents the voltage measured across the reference capacitor of the Sawyer-Tower circuit. In this case, electrical breakdown inside the film has not yet occurred; (b) the triangle waveform is still the input voltage signal but the voltage measured across the reference capacitor jumps beyond the voltage limit of the oscilloscope, which means that electrical breakdown has occurred.}
\vspace{-0.15in}
\end{center}
\end{figure}

Let's assume we apply an ac voltage signal with an amplitude $V$ and a frequency $f$ to our polymer film. The time interval between starting to apply the voltage to the film and the occurrence of electrical breakdown in the film is $T_{D}$. It is then easy to determine the value of $N_{f}$ that is $N_{f}=f\times T_{D}$. The value of $P_{r}$ (the last recorded remnant polarization before the occurrence of electrical breakdown) can also be directly obtained from the measured E-P hysteresis loops (here we only consider the positive values of $P_{r}$). Both $P_{f}$ and $\beta_{CM}$ cannot be directly measured. We took advantage of a physical concept, which has been used to solve certain problems in continuous phase transitions \cite{fu2013}, the universal dielectric relaxation \cite{fu2014}, and the electric-field-induced elastic fatigue in poly(vinylidene fluoride) (PVDF) \cite{he2014}, and a curve fitting formula that is related to the Weibull distribution \cite{rinne2008} to indirectly determine $P_{f}$ and $\beta_{CM}$. Our experimental data, curve fitting of measurements, comments, and discussions are given in the next section.

\section{\label{sec:results}Results and discussions}
In our experiment, a series of triangle voltage waveforms with a frequency $f=10$Hz and an amplitude $V$ ranging from $V_{min}=1.3$MV/cm to $V_{max}=1.7$MV/cm were used to induce polarization fatigue failure in the polymer film. For each applied signal, by using the method mentioned in Section~\ref{sec:material}, we measured $P_{r}$ and $N_{f}$. We also measured the remnant polarization, $P_{r}^{1}$, of the first reversal under the waveform with the amplitude of $V_{max}=1.7$MV/cm and then divided all measured $P_{r}$ by $P_{r}^{1}$; thus, in the $N_{f}-P_{r}$ relation given in this section, $P_{r}$ is always normalized. All pairs of $N_{f}$ and $P_{r}$ are marked in Fig. 4. We now need to find what kind of fitting function can link all $N_{f}-P_{r}$ pairs together. Since the values of $P_{r}$ are normalized, we propose that the following fitting function might have the best fit to those $N_{f}-P_{r}$ pairs. The physics behind this function will be explained after the curve fitting result is obtained.
\begin{equation}
P_{r}=q_{c}N_{f}^{(k_{w}-1+\phi)}\mathrm{exp}\left[-\left(\frac{N_{f}}{N_{w}}\right)^{k_{w}}\right],
\label{pdfweibull}
\end{equation}
here, for convenience, $k_{w}$ and $N_{w}$ are called the shape parameter and the scale parameter, respectively; since $P_{r}$ is normalized, we can estimate $q_{c}\sim1$; the meanings of both $q_{c}$ and $\phi$ will be explained later.

After considerable computational effort, we found that Eq.~(\ref{pdfweibull}) does have the best fit to those $N_{f}-P_{r}$ pairs if
we adopted the following parameters: $q_{c}=1.07$, $k_{w}=0.53$, $\phi=0.462$, $N_{w}=3.6\times10^{4}$. The fitting result is given in Fig. 4. Comparing Eq.~(\ref{pdfweibull}) to the generalized Coffin-Manson law, Eq.~(\ref{fitting-function}), we obtained the Coffin-Manson exponent, $\beta_{CM}=1-k_{w}-\phi=0.008$, and the electrical failure coefficient, $P_{f}$, which is a natural exponential function given below.
\begin{eqnarray}
P_{f} & = & q_{c}\mathrm{exp}\left[-\left(\frac{N_{f}}{N_{w}}\right)^{k_{w}}\right] \nonumber \\
& = & 1.07\mathrm{exp}\left[-\left(\frac{N_{f}}{3.6\times10^{4}}\right)^{0.53}\right]
\label{failurecoefficient}
\end{eqnarray}
The above formula does not surprise us since we know that $P_{f}$ must be a time-dependent parameter. Therefore, the generalized Coffin-Manson law for electrical fatigue failure of our polymer films is:
\begin{equation}
P_{r}\approx P_{f}(N_{f})^{-0.008}
\label{f-f1}
\end{equation}
\begin{figure}[h!]
\begin{center}
\includegraphics[width=1.0\columnwidth]{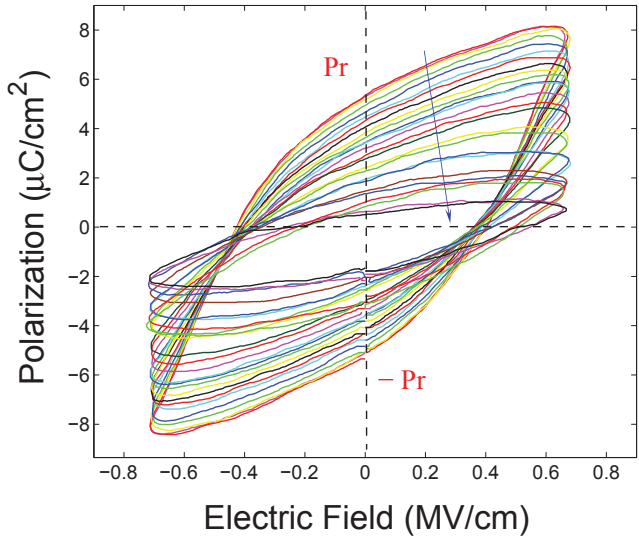}
\vspace{-0.10in}
\caption{Remnant polarization fatigue observed in the P(VDF-TrFE) copolymer film shown in Fig. 1.}
\vspace{-0.05in}
\end{center}
\end{figure}

So why does Eq.~(\ref{f-f1}) have the perfect fit to those $N_{f}-P_{r}$ pairs? To answer this question, we should start from the Weibull distribution, which is written below.
\begin{equation}
f(x;\lambda,k)=\left\{\begin{array}{ll}
                \frac{k}{\lambda^{k}}x^{k-1}\mathrm{exp}\left[-\left(\frac{x}{\lambda}\right)^{k}\right] & \ \ \ \ \ \mbox{$x\geq0$} \\
                0 & \ \ \ \ \ \mbox{$x<0$}
                \label{weibulldistribution}
                \end{array}
\right.
\end{equation}
where $f(x;\lambda,k)$ and $x$ represent a probability density function and a random variable, respectively; $k>0$ is the shape parameter and $\lambda>0$ is the scale parameter. The Weibull distribution is a continuous probability distribution and has a broad range of applications in many fields. One of them is to describe a particle size distribution \cite{rinne2008}. To be specific, we consider a fluid; there are particles dispersed in it; the Weibull distribution is a mathematical function that gives the fraction of particles that present according to size. In this study, we only care about the case of $x\geq0$ in the above equation; we replace $f(x;\lambda,k)$ with $P_{r}$, $x$ with $N_{f}$, $\lambda$ with $N_{w}$, $k$ with $k_{w}$, $\frac{k}{\lambda^{k}}$ with $q_{c}$ ($q_{c}$ corresponds to the normalized values of a series of $P_{r}$, while, in Eq.~(\ref{weibulldistribution}), $\frac{k}{\lambda^{k}}$ does not correspond to the normalized values of a series of $f(x;\lambda,k)$. So the substitution of $q_{c}$ for $\frac{k}{\lambda^{k}}$ means that $f(x;\lambda,k)$ has already been normalized here.); add one more parameter, $\phi$, to the exponent of $N_{f}$. Then we can convert Eq.~(\ref{weibulldistribution}) of the spatial domain into Eq.~(\ref{pdfweibull}) or the generalized Coffin-Manson law defined by Eq.~(\ref{fitting-function}) of the time domain. The reason we did such a conversion is explained below.

Both mechanical and electrical fatigue phenomena reflect the nature of irreversible physical processes, in which the crystalline phase at low energy level is repeatedly perturbed by the applied cyclic loading and parts of them will be excited to become the disordered atomic phase at high energy level. Disordered structures in that high energy atomic phase continuously grow and aggregate during the cyclic loading; at some point, the chemical bonds holding those atoms of disordered structures will start to be partially broken and microcracks (the term adopted for mechanical fatigue failure) or partial discharge channels (the term adopted for electrical fatigue failure or electrical breakdown) will soon emerge. These ``{\it structure defects}" will significantly reduce the value of $W_{max}$ at certain local points where they aggregate. When the material cannot withstand the strength of the cyclic loading at those local spots, fatigue failure occurs. Clearly, the occurrence of fatigue failure depends on both the loading and the amount of disordered structures. Since disordered structures grow and aggregate during the cyclic loading, a large amount of disordered structures always mean the presence of more large-size disordered structures, which also mean the presence of more microcracks or partial discharge channels. We now re-examine the $N_{f}-P_{r}$ relation shown in Fig. 4; we can find that what $N_{f}-P_{r}$ actually represents is a disordered-structure size distribution: on the left side, failure can occur in the presence of less large-size disordered structures or partial discharge channels due to large electrical loadings; on the right side, failure can only occur after more large-size disordered structures or partial discharge channels aggregate due to small electrical loadings. This is the reason that we propose to use Eq.~(\ref{pdfweibull}) to fit those $N_{f}-P_{r}$ pairs.

\begin{figure}[h!]
\begin{center}
\includegraphics[width=1.0\columnwidth]{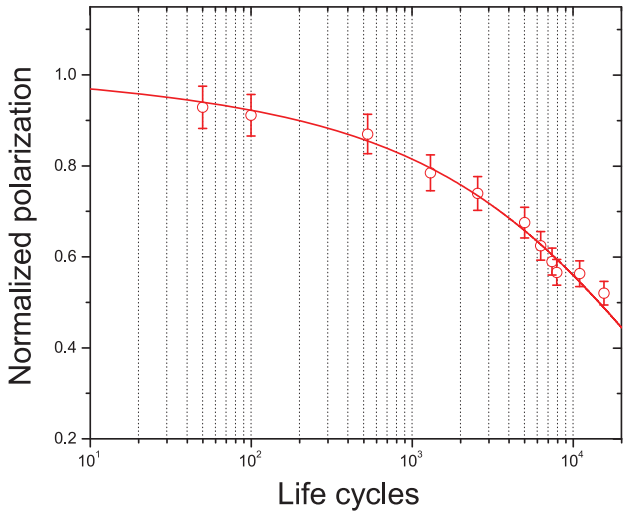}
\vspace{-0.10in}
\caption{The measured remnant polarization ($P_{r}$) and the number of life cycles to failure ($N_{f}$) are shown here; the curve represents the fitting result by using the Weibull distribution.}
\vspace{-0.05in}
\end{center}
\end{figure}

Now we have to answer the question why the fatigue failure phenomena are governed by a class of particle size distribution? Recent theoretical studies have shown that those disordered structures form a nematic phase or a partially ordered liquid phase \cite{fu2013,fu2014}. Without external perturbations, such a nematic phase belongs to the $\mathrm{D_{\infty h}}$ point-group and does not show polarity \cite{blinov2010,degennes1993}. However, under external perturbations, the $\mathrm{D_{\infty h}}$ symmetry is broken and a polar order will emerge in this ``{\it modified}" nematic phase that can be regarded as an exotic nematic phase \cite{degennes1993}, in which its point-group symmetry becomes $\mathrm{C_{\infty v}}$. So how does the nematic phase change under an external perturbation? According to Le Chatelier's principle \cite{landau1980}, the nematic phase would undergo a specific structural change to counteract any imposed change by the perturbation. If an electrical cyclic loading voltage is applied to a piece of ferroelectric material, the nematic phase will undergo a change to counteract the spontaneous polarization (the induced remnant polarization when the amplitude of the voltage reaches zero in our case) inside the material (see Ref.~\cite{fu2014} for a detailed explanation); Thus, the polar components of the exotic nematic phase must counteract the induced remnant polarization. Obviously, the larger the volume of the exotic nematic phase (the amount of disordered structures) presents, the severer fatigue or the easier failure occurs. Therefore, fatigue failure is always linked to the size distribution of those disordered structures.

For the polymer film fabricated on a substrate, the distribution of the nematic phase is not uniform; the major part of this phase concentrates in the top layers lying beneath the surface. The formation of the nematic phase in these top layers needs to overcome the atomic binding energy, which mainly comes from the chemical bond and the intermolecular force of the film; however, the formation of the same phase in the bottom layers near the interface between the film and the substrate needs to overcome not only that binding energy but also the clamping energy, which comes from the atomic clamping force between the film and the substrate. Thus, the size of the nematic phase in the top layers would be much larger than that in the bottom layers. This is the reason that an asymmetric remnant polarization fatigue process can be seen in Fig. 3. In fact, such asymmetric polarization fatigue can also be seen in both Figs. 1 of Refs.~\cite{zhu2006,yuan2011}. Only is the degree of asymmetry in the fatigue processes they observed not as significant as that we observed is. This ``{\it difference}" can be explained as follows. In Ref.~\cite{zhu2006}, for instance, their film has a thickness of $\sim\mathrm{1\mu m}$ and was annealed at 145$^{\circ}$C for 6 hours. As comparison, our film has a thickness of $\sim\mathrm{3\mu m}$ and was not annealed. The smaller the thickness of film is, the smaller the influence of that ``{\it clamping effect}" has. The fraction of the crystalline phase in their film should be much greater than that of ours (because their film was annealed). This means that the fraction of the nematic phase in their film should be much less than that in ours. This is why the asymmetric phenomenon in the polarization fatigue processes of their films is not easy to be observed.

The parameter, $\phi$, given in Eq.~(\ref{pdfweibull}), is the factor that represents the contribution of amorphous phases to the measured remnant polarization of our P(VDF-TrFE) films. Without external perturbations, amorphous phases of P(VDF-TrFE) films usually do not show polarity. However, under a strong electric field, a field-induced phase transition can occur in amorphous phases and generate extra polar components. Such a process was not considered when the conventional Weibull distribution function was developed. Thus, we have to add this parameter to the exponent of $N_{f}$ in Eq.~(\ref{pdfweibull}).

Finally, it might be worthwhile to point out one of interesting features of the Coffin-Manson exponent. In the past several decades, it has been observed that the Coffin-Manson exponent possesses the remarkable universality; $\beta_{CM}\sim0.5$ has been found in many single-phased metallic materials \cite{coffin-manson1992}. Since P(VDF-TrFE) is viscoelastic in our case, we cannot directly compare our measured $\beta_{CM}$ to the Coffin-Manson exponent of metallic materials. If we define a generalized Coffin-Manson exponent by considering the influence of amorphous phases on the polarization fatigue of P(VDF-TrFE) films as $\widetilde{\beta}_{CM}=\beta_{CM}+\phi=1-k_{w}$, then we obtain $\widetilde{\beta}_{CM}=0.47$ from our experimental data. Are Coffin-Manson exponents measured in different materials (for instance, single-phased metallic materials and P(VDF-TrFE) copolymers) belonging to the same universality class? Our current experimental studies cannot provide a definitive answer. Such unusual universality may only be explained by using statistical mechanics. Unfortunately, no convincing explanation has been given yet. It is our wish that the experimental studies reported in this letter could provide a different perspective to revisit this problem.

\section{Concluding remarks}
Our experimental studies have shown that a generalized Coffin-Manson law has the best fit to the electrical fatigue failure data collected from P(VDF-TrFE) copolymer films. We also found that both the conventional and generalized Coffin-Manson laws actually belong to a class of the two-parameter Weibull distribution function. It indicates that the formation and evolution of disordered structures in both cases are very similar at the atomic level. Therefore, we conjecture that both mechanical and electrical fatigue and failure are governed by the same physical law.

\section*{Acknowledgments}
Shaopeng Pei kindly provided experimental assistance at the early stage of this work. His help is greatly appreciated.
%

%
%

%

%
%
%
\end{document}